\documentclass[preprint,showpacs,preprintnumbers,amsmath,amssymb]{revtex4}

\usepackage{graphicx}
\usepackage{dcolumn}
\usepackage{bm}

\usepackage{feynmf}
\usepackage{slashed}

\begin{document}

\title{One Right-handed Neutrino to Generate Complete Neutrino Mass Spectrum in the Framework of NMSSM}
\author{ Yi-Lei Tang}

\affiliation
{Institute of Theoretical Physics, Chinese Academy of Sciences, \\
and State Key Laboratory of Theoretical Physics,\\
P. O. Box 2735, Beijing 100190, China}

\email{tangyilei10@itp.ac.cn}

\date{\today}

\begin{abstract}

The see-saw mechanism is usually applied to explain the lightness of neutrinos. The traditional see-saw mechanism introduces at least two right-handed neutrinos for the realistic neutrino spectrum. In the case of supersymmetry, loop corrections can also contribute to neutrino masses, which lead to the possibility to generate the neutrino spectrum by introducing just one right-handed neutrino. To be realistic, MSSM suffers from the $\mu$ problem and other phenomenological difficulties, so we extend NMSSM (the MSSM with a singlet S) by introducing one single right-handed neutrino superfield (N) and relevant phenomenology is discussed.

\end{abstract}


\maketitle
\section{Introduction}

Up to now, experiments have established the fact that neutrinos oscillate among each other. Two mass-squared differences ($\delta m_{12}^2$, $\delta m_{23}^2$), together with all three mixing angles($\theta_{12}$, $\theta_{23}$, $\theta_{13}$) have been measured\cite{2012te}, prompting us that at least two generations of neutrinos among the three are massive. Type I see-saw Mechanism is a way to generate small Majorana neutrino masses\cite{SS} by introducing lepton number violating $\Delta L=2$ Majorana mass term for right-handed neutrinos. On the other hand, in the framework of supersymmetry, after the lepton number is violated, one-loop radiative corrections naturally generate non-zero neutrino mass terms\cite{AZee}.

It is possible to generate realistic neutrino spectrum by introducing just one right-handed neutrino in supersymmetry. Minimal supersymmetry standard model (MSSM) extended with one right-handed neutrino is discussed in \cite{OneRightHandedMSSM}. Considering both tree-level see-saw mechanism and the one-loop radiative corrections\cite{AZee}, one right-handed neutrino is enough to generate masses for three generations of neutrinos. However, MSSM suffers from the $\mu$ problem. Next to minimal supersymmetry standard model (NMSSM)\cite{NMSSMInt} extended with one right-handed neutrino was originally discussed in \cite{TriBiNMSSM},  \cite{MultiRightHandedNMSSM} \cite{RHSDarkMatter}, where right-handed neutrinos acquire TeV-scale Majorana mass terms through their couplings with the singlet Higgs superfield S introduced in NMSSM. This coupling also establishes the connection between the right-handed neutrinos with the Higgs sectors, which may influence the phenomenology of the Higgs bosons. The possibility that right-handed neutrino in the framework of NMSSM may contribute to the Higgs boson mass is also discussed in \cite{LotsofMistakes}.

In this paper, we will show that NMSSM extended with a single right-handed neutrino superfield can generate the complete neutrino mass spectrum considering both contributions from the tree-level Type-I see-saw mechanism and one-loop radiative corrections. We will show that this model contains all the possibilities of the size of mixing angles. It is because once the correct mass spectrum is generated, it is almost free for us to rotate the mass matrix, with little experimental limitations. We also considered other experimental constraints and calculated the corrections of the Higgs Boson mass from the contributions of the right-handed neutrino\cite{LotsofMistakes}, and gave some numerical results.

\section{Model and Calculation}

Here we impose a global $Z_3$ symmetry and keep the R-parity conservation, just as usual NMSSM with $Z_3$ symmetry\cite{NMSSMInt}. The $Z_3$ quantum number assigned to the right-handed neutrino $N$ is the same as Singlet Higgs $S$, while the R-parity of $N$ is set as positive, just as other MSSM matter superfields. Thus the involved superpotential is strongly limited into the form
\begin{eqnarray}
W_{part} = \lambda_N S N N + y_N H_u \cdot L N + \lambda S H_u \cdot H_d + \frac{\kappa}{3} S^3, \label{Superfield}
\end{eqnarray}
where $S$ is the singlet Higgs superfield originally existed in NMSSM, and $N$ is the right-handed neutrino. Here we only listed the terms involving lepton and Higgs fields.

The relevant soft terms breaking the supersymmetry are listed below,
\begin{eqnarray}
-\mathcal{L}_{soft} &\supset& M_{H_u}^2 |H_u|^2 + M_{H_d}^2 |H_d|^2 + M_s^2 |S|^2 + (\lambda A_\lambda H_u \cdot H_d S + \frac{\kappa}{3} A_{\kappa} S^3 + h.c. ) + m^{l2}_{ij} \tilde{L}_i^\dagger \tilde{L}_j \nonumber \\
&& + M_{\tilde{N}}^2 |\tilde{N}|^2 + (\lambda_N A_N S \tilde{N} \tilde{N} + h.c. ) + (y_{N} A_{yN} H_u \tilde{L} \tilde{N} + h.c.)  \label{Soft_Term}
\end{eqnarray}
where $m^{l2}_{ij}=m^{l2}_{ji}$.

One might consider a more compact model that $N$ just plays the role of $S$ \cite{munuMSSM}. However, this model breaks R-parity and do not contain dark matter.

If scalar $\tilde{S}$ acquires a vacuum expectation value(vev) $v_s$,
\begin{eqnarray}
S=v_s + \frac{S_R + i S_I}{\sqrt{2}},
\end{eqnarray}
terms like $\tilde{N}\tilde{N}$ appears, which supplies the $\Delta L=2$ quadratic terms in the sneutrino sector, contributing to the one-loop neutrino mass corrections.

The vevs of the doublet Higgs fields are defined as
\begin{eqnarray}
H_u^0 = v_u + \frac{Re(H_u^0) + i Im(H_u^0)}{\sqrt{2}},
\text{\qquad}
H_d^0 = v_d + \frac{Re(H_d^0) + i Im(H_d^0)}{\sqrt{2}}
\end{eqnarray}

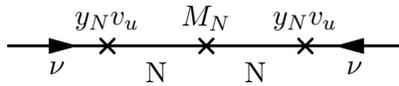
\begin{figure}
\centering
\begin{fmffile}{treelevel}
\begin{fmfgraph*}(150,60)
    \fmfleft{vin}
    \fmfright{vout}
    \fmfv{decoration.shape=cross, decoration.size=60, label=$y_N v_u$, label.angle=90}{v1}
    \fmfv{decoration.shape=cross, decoration.size=60, label=$M_N$, label.angle=90}{v2}
    \fmfv{decoration.shape=cross, decoration.size=60, label=$y_N v_u$, label.angle=90}{v3}
    \fmf{fermion,label=$\nu$}{vin,v1}
    \fmf{plain,label=N}{v1,v2}
    \fmf{plain,label=N}{v2,v3}
    \fmf{fermion,label=$\nu$, label.side=left}{vout,v3}    
\end{fmfgraph*}
\end{fmffile}
\caption{Tree-level see-saw neutrino mass.} \label{Tree_See_Saw}
\end{figure}

Neutrinos then acquire tree-level Majorana mass terms after integrating out the fermionic $N$ through Figure \ref{Tree_See_Saw} \cite{LotsofMistakes},
\begin{eqnarray}
M_{\nu i j}^{TreeLevel} = -y_{Ni} y_{Nj} \frac{v_u^2}{M_N}.
\end{eqnarray}
Because $rank(\left\lbrace y_{Ni} y_{Nj} \right\rbrace) = 1$, matrix $\{M_{\nu i j}^{TreeLevel}\}$ has only one non-zero eigenvalue, which leave other neutrinos massless.

\begin{figure}
\centering
\begin{fmffile}{looplevel}
\begin{fmfgraph*}(150,60)
    \fmfbottom{nu1,nu2}
    \fmfpen{thick}
    \fmf{fermion,label=$\nu$}{nu1,v1}
    \fmf{dashes,left=0.2,tension=1,label=$\tilde{v}_i$,label.side=right}{v1,v2}
    \fmf{dashes,left=0.2,tension=1,label=$\tilde{v}_j$,label.side=right}{v2,v3}
    \fmf{dashes,left=0.2,tension=1,label=$\tilde{N}$,label.side=right}{v3,v4}
    \fmf{dashes,left=0.2,tension=1,label=$\tilde{N}$,label.side=right}{v4,v5}
    \fmf{dashes,left=0.2,tension=1,label=$\tilde{v}_i$,label.side=right}{v5,v6}
    \fmf{plain,label=$\tilde{\chi}$}{v1,v6}
    \fmf{fermion,label=$\nu$,label.side=left}{nu2,v6}
    \fmffreeze
    \fmfshift{-0.15w,0h}{v1}
    \fmfshift{0.15w,0h}{v6}
    \fmfshift{-0.14w,.5h}{v2}
    \fmfshift{.14w,.5h}{v5}
    \fmfshift{-0.07w,.72h}{v3}
    \fmfshift{0.07w,.72h}{v4}
    \fmfv{decoration.shape=cross, decoration.size=60, label=$m_{\nu ij}$, label.angle=180}{v2}
    \fmfv{decoration.shape=cross, decoration.size=60, label=$y_N v_u A_{yN}$, label.angle=125}{v3}
    \fmfv{decoration.shape=cross, decoration.size=60, label=$M_{\tilde{N}\tilde{N}}^2$, label.angle=55}{v4}
    \fmfv{decoration.shape=cross, decoration.size=60, label=$y_N v_u A_{yN}$, label.angle=0}{v5}
\end{fmfgraph*}
\end{fmffile}
\caption{One-loop diagram to generate neutrino mass by ``Mass Insertion Method''.} \label{Loop_Level}
\end{figure}
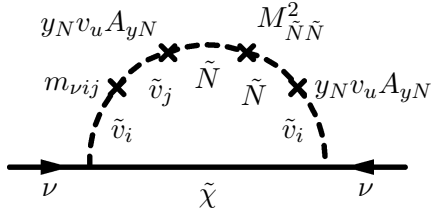

Now that we are considering a supersymmetric theory, each particle is paired up with a super-partner, so is the right-handed neutrino. Thus, right-handed scalar-neutrino contribute into the mass terms of neutrino through radiative corrections in Figure \ref{Loop_Level}.  As mentioned in \cite{OneRightHandedMSSM}, in order for the loop-level neutrino mass terms not to be aligned with the tree level ones, we also need cross terms in the Sneutrino soft mass-squared matrix $m_{\nu i j}$ which result in flavor-changing neutral current (FCNC) processes in the lepton sector.

From Figure \ref{Loop_Level}, we can estimate the loop-level neutrino mass,
\begin{eqnarray}
m_{oneloop} \sim \left( \text{Loop Factor} \right) m^{l2}_{i j} v_u^2 g^2 \frac{y_N y_N A_{yN} A_{yN} M_{\tilde{N} \tilde{N}}^2}{ M^7 },
\end{eqnarray}
where M is the typical mass scale of the propagators in the loop, $m^{l2}_{ i j}$ is the off-diagonal elements of the soft mass-square matrix of the left-handed leptons. Fortunately, for TeV see-saw mechanics, $y_N$ tend to be rather small and are $\sim 10^{-6}$, which is comparable with the electron Higgs Yukawa coupling term, thus allows relatively large $A_{yN}$, which significantly increase the $m_{oneloop}$, and its phenomenological effects are also highly suppressed by the factor $y_N$ which always appear together with $A_{yN}$.

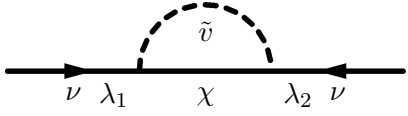
\begin{figure}
\centering
\begin{fmffile}{looplevelMassEigen}
\begin{fmfgraph*}(150,60)
    \fmfbottom{nu1,nu2}
    \fmfpen{thick}
    \fmf{fermion,label=$\nu$}{nu1,v1}
    \fmf{dashes,left=1,tension=0,label=$\tilde{v}$,label.side=right}{v1,v2}
    \fmf{fermion,label=$\nu$,label.side=left}{nu2,v2}
    \fmf{plain,label=$\chi$}{v1,v2}
    \fmfv{label=$\lambda_1$}{v1}
    \fmfv{label=$\lambda_2$}{v2}
\end{fmfgraph*}
\end{fmffile}
\caption{one-loop neutrino mass calculated under mass eigen-state basis} \label{Loop_Level_Mass_Basis}
\end{figure}

Figure \ref{Loop_Level} is based upon the ``Mass Insertion Method'', which is clear in concept, however is complicated to be calculated when the number of ``crosses'' inserted into a propagator are many. Unlike \cite{OneRightHandedMSSM}, in this paper we only use ``Mass Insertion Method'' to analyse qualitatively however calculate directly in mass-eigenstate basis quantitatively through Figure \ref{Loop_Level_Mass_Basis}.

Calculating in mass-eigenstate basis according to Figure \ref{Loop_Level_Mass_Basis} involves a summation over a group of graphs with different real scalar propagators. Each graph in Figure \ref{Loop_Level_Mass_Basis} is infinite, unlike Figure \ref{Loop_Level}. If we use the dimensional regularization scheme, all $\frac{1}{\epsilon}$ terms appear in each diagram should be accurately cancelled, the non-zero remains of the finite part is just due to difference of the masses and the mixing between the real part and the imaginary part of each scalar field. The mixing among scalar fields cannot be omitted even though they're small, exceeding the capability of ordinary computational numeric float-point data types, however, we use gmp/mpfr to deal with it.

Calculating one single diagram in Figure \ref{Loop_Level_Mass_Basis} generates the result
\begin{eqnarray}
m_{OneDiagram}=\frac{\lambda_1 \lambda_2 m_f}{4 \pi^2} \frac{m_f^2 - m_s^2 + m_f^2 \ln{\frac{\mu^2}{m_f^2}} - m_s^2 \ln{\frac{\mu^2}{m_k^2}}}{m_f^2-m_s^2}, \label{Single_Diagram_1}
\end{eqnarray}
particularly, when $m_s \rightarrow m_f$,
\begin{eqnarray}
m_{OneDiagram}=\frac{\lambda_1 \lambda_2 m_f}{4 \pi^2} \ln{\frac{\mu^2}{m_f^2}}, \label{Single_Diagram_2}
\end{eqnarray}
where $m_f$, $m_s$ separately indicates the mass of the Majorana particle and the real scalar particle running in the loop. $\mu$ can be any mass-scale and must be accurately cancelled into disappearance after summing over all corresponding diagrams. In (\ref{Single_Diagram_1}) and (\ref{Single_Diagram_2}), we also dropped the divergent $\frac{1}{\epsilon}$ for simplicity, which we know that should also be cancelled finally. 

We define
\begin{eqnarray}
\tilde{N}=\frac{\tilde{N}_R + i \tilde{N}_I}{\sqrt{2}}, \text{\qquad}
\tilde{\nu}_i = \frac{\tilde{\nu}_{iR} + i \tilde{\nu}_{iI}}{\sqrt{2}}.
\end{eqnarray}
Notice that according to (\ref{Superfield}) (\ref{Soft_Term}), when $\tilde{S}$ acquires vacuum expectation value $v_s$, $\tilde{N} \tilde{N}$ terms are generated and thus split the mass spectrum of $\tilde{N}_R$ and $\tilde{N}_I$, and influence the spectrum of $\tilde{\nu}_R$ and $\tilde{\nu}_I$ through mixture between right-handed and left-handed sneutrinos. The result of the $8 \times 8$ mass-square matrix of sneutrinos is showed below,
\begin{eqnarray}
  \mathcal{L} \supset - \left[
  \begin{array}{cccc}
    \tilde{v}_{iR}^* &  \tilde{N}_R^* & \tilde{v}_{iI}^* & \tilde{N}_I^*
  \end{array} \right]
  \left[
  \begin{array}{cccc}
    M_{\nu, 3 \times 3} & y_N A_{yN} v_u & 0 & 0 \\
    y_N A_{yN} v_u & M_R^2 & 0 & 0 \\
    0 & 0 & M_{\nu, 3 \times 3} & y_N A_{yN} v_u \\
    0 & 0 & y_N A_{yN} v_u & M_I^2
  \end{array} \right]
  \left[
  \begin{array}{c}
    \tilde{v}_{iR} \\
    \tilde{N}_R \\
    \tilde{v}_{iI} \\
    \tilde{N}_I
  \end{array} \right], \label{Sneutrino_Mass_Matrix}
\end{eqnarray}
where $M_{\nu 3 \times 3}$ is just the ordinary mass matrix of sneutrinos in NMSSM,
\begin{eqnarray}
M_{\nu 3 \times 3} = [m^{l2}_{ij}] + \frac{1}{2} m_Z^2 \cos{2 \beta},
\end{eqnarray}
where $[m^l_{ij}]$ is the supersymmetry breaking soft mass matrix of left-handed leptons. In addition,
\begin{eqnarray}
M_R^2 = 4 \lambda_N^2 v_s^2 + M_{\tilde{N}}^2 + 2 \lambda_N v_s A_N + 2 \lambda_N ( \kappa v_s^2 - \lambda v_u v_d ) \nonumber \\
M_I^2 = 4 \lambda_N^2 v_s^2 + M_{\tilde{N}}^2 - 2 \lambda_N v_s A_N - 2 \lambda_N ( \kappa v_s^2 - \lambda v_u v_d ) \label{R_I_Mass}
\end{eqnarray}

From observing (\ref{Sneutrino_Mass_Matrix}) together with (\ref{R_I_Mass}), we can learn that the mass split of the real and the imaginary part of $\tilde{N}$ was transferred into $\tilde{\nu}_i$ by the mixing term $y_N A_{yN} v_u$. Then, there are 8 different real-scalar sneutrinos out of 3 left-handed neutrino and one right-handed neutrino superfields.

Diagram \ref{Loop_Level_Mass_Basis} involves the neutrilinos. There are 5 neutrilinos in NMSSM theory, each is a mixture of bino, neutral wino, 2 higgisinos and one singlino. sneutrinos interact with the neutral gauginos through the supersymmetry electro-weak gauge vertices, while interact with the higgisinos through the Yukawa vertices. Left-handed neutrinos do not directly interact with singlinos. However, on the circumstances of the Tev-scale see-saw mechanism, gauge coupling constant (typically $\sim 0.3$) is much greater than Yukawa coupling constant (typically $\sim 10^{-7} - 10^{-8}$), so the radiative one-loop contribution is mainly due to the bino and neutral winos. The $\tilde{\nu} \tilde{\nu} \chi$ type coupling constant matrix for each neutrilino in the basis of $[\tilde{B}~~\tilde{W}^0~~\tilde{H}_u^0]$ is showed below,
\begin{eqnarray}
\mathcal{L} \supset & & \bar{\nu}_i \frac{1-\gamma^5}{2} B \left[
\begin{array}{cccccccc}
-g_1 & 0 & 0 & 0 & i g_1 & 0 & 0 & 0 \\
0 & -g_1 & 0 & 0 & 0 & i g_1 & 0 & 0 \\
0 & 0 & -g_1 & 0 & 0 & 0 & i g_1 & 0
\end{array} \right]_{ij} \tilde{\nu}_j^\prime \nonumber \\
 &+& \bar{\nu}_i \frac{1-\gamma^5}{2} W^0 \left[
\begin{array}{cccccccc}
-g_2 & 0 & 0 & 0 & i g_2 & 0 & 0 & 0 \\
0 & -g_2 & 0 & 0 & 0 & i g_2 & 0 & 0 \\
0 & 0 & -g_2 & 0 & 0 & 0 & i g_2 & 0
\end{array} \right]_{ij} \tilde{\nu}_j^\prime \label{Loop_Needed_Interaction} \\
&+& \bar{\nu_i} \frac{1-\gamma^5}{2} H_u^0 \frac{1}{2 \sqrt{2}}\left[
\begin{array}{cccccccc}
0 & 0 & 0 & y_{N1} & 0 & 0 & 0 & i y_{N1} \\
0 & 0 & 0 & y_{N2} & 0 & 0 & 0 & i y_{N2} \\
0 & 0 & 0 & y_{N3} & 0 & 0 & 0 & i y_{N3}
\end{array} \right]_{ij} \tilde{\nu}_j^\prime, \nonumber
\end{eqnarray}
where
\begin{eqnarray}
\tilde{\nu}_j^\prime = \left[
\begin{array}{cccc}
\tilde{\nu}_{iR} & \tilde{N}_R  & \tilde{\nu}_{iI} & \tilde{N}_I
\end{array} \right]_i .
\end{eqnarray}

In numerical calculation, we rotate (\ref{Loop_Level_Mass_Basis}) into mass eigenstate by multiplying the matrices with the neutrilino transforming matrix supplied by NMSSMtools, and the sneutrino transforming matrix is calculated according to (\ref{Sneutrino_Mass_Matrix}).

We have to mention that this model do not have the ability to predict any mixing angles, that is to say, any mixing angle is permitted if only the correct mass-squared difference is acquired. If we get an example of neutrino mass matrix $M_\nu$ with the correct mass spectrum, then we can always find a unitary matrix $V$ to rotate $M_\nu$ into the ``correct'' matrix $M_\nu^{correct}$ with the ``correct'' mixing angles, that is to say,
\begin{eqnarray}
M_\nu \rightarrow M_\nu^{correct} = V^* M_\nu^\prime V^\dagger.
\end{eqnarray}
Use the same $V$ to operate the all of the sneutrino soft mass-square matrix, the $H_u L_i N$ Yukawa coupling constants $y_{Ni}$, and the A-terms $A_{yNi} y_{Ni}$ altogether into a new group of parameters to input into the theory,
\begin{eqnarray}
m^{l2} \rightarrow m^{l2}_{correct} = V m^{l2} V^\dagger \nonumber \\
y_N \rightarrow V^\dagger y_N \\
y_N A_{yN} \rightarrow V^\dagger y_N A_{yN} \nonumber .
\end{eqnarray}
Then we can always acquire ``correct'' $M_\nu^\prime$. What we only need to consider is that whether these operations involving $m^{l2}_{ij}$ disturb the off-diagonal terms which generate large FCNC. In fact, as $|V_{ij}| \leq 1$ always hold, so the magnitude of $m_{ij}(i \neq j)$ do not change. Therefore, in the processes of numeric calculation, we only concern about the neutrino mass-square hierarchy.

\section{Numerical Results and Analyse}

We modified NMSSMTools-4.2.1\cite{NMSSMTools1}\cite{NMSSMTools2} by adding the effects of our extended sectors. The $H_u L N$ Yukawa coupling constants and the corresponding A terms are so small that their phenomenology effects are highly suppressed, thus we needn't consider them. We checked and followed \cite{LotsofMistakes} together with \cite{NMSSMHiggs} to calculate the loop contribution to Higgs mass, and see appendix for the process and formulae in detail. We also assume that only R-parity odd $\tilde{N}$ can be the candidate of dark matter, thus we modified the model files contained in the MicroOMEGAS\cite{MicroOMEGAS} inside the NMSSMTools.

We opened the constraint on muon anomalous $g_\mu-2$ when scanning, which is so sensitive to the supersymmetry breaking soft masses of sleptons. Though we need non-zero off-diagonal elements of the soft mass-square matrix of the left-handed leptons, they are constrained by experimental limits such as the branching ratio of $\mu \rightarrow e + \gamma$\cite{primer}. In order to avoid such constraints, we need either relatively large slepton masses or small off-diagonal terms. However, the constraint of muon's $g_\mu -2$ prefers smaller supersymmetry breaking soft terms of sleptons, so we set the range $(500GeV)^2 < m^l_{11} = m^l_{22} = m^l_{33} < (1500GeV)^2$, and $m^l_{ij}<\frac{1}{10} m^l_{ii}$ for each $i \neq j$. This scale cannot avoid FCNC $\mu \rightarrow e + \gamma $ completely, however, which will be discussed in the next section.

When calculating neutrino mass matrix, cases are that accuracy of beyond $10^{-20}$ is needed, so we used the numerical tools gmp/mpfr. However, this technique extremely slows down the speed, so we scanned avoiding to consider the neutrino masses at first, then calculate the neutrino mass matrices by testing in $y_{N}$-$A_{yN}$-$m^l_{ij}(i\neq j)$ parameter space for each parameter point passed the previous constraints. We should note that if the lepton-number $U(1)_L$ symmetry does not break, diagrams in Fig. \ref{Loop_Level_Mass_Basis} cancel with each other strictly. It is due to the existence of the lepton-number violating terms $\lambda_N S N N$ and $\lambda_N A_N S \tilde{N} \tilde{N}$ that different diagrams in Fig. \ref{Loop_Level_Mass_Basis} cannot be cancelled out strictly, leaving a small finite value, looking as if we are tuning something. As mentioned before, we needn't care about the mixing angles as they can always be acquired after exerting the mentioned process on each parameter point without disturbing the phenomenology.

\begin{figure}[center]
\includegraphics[width=4in]{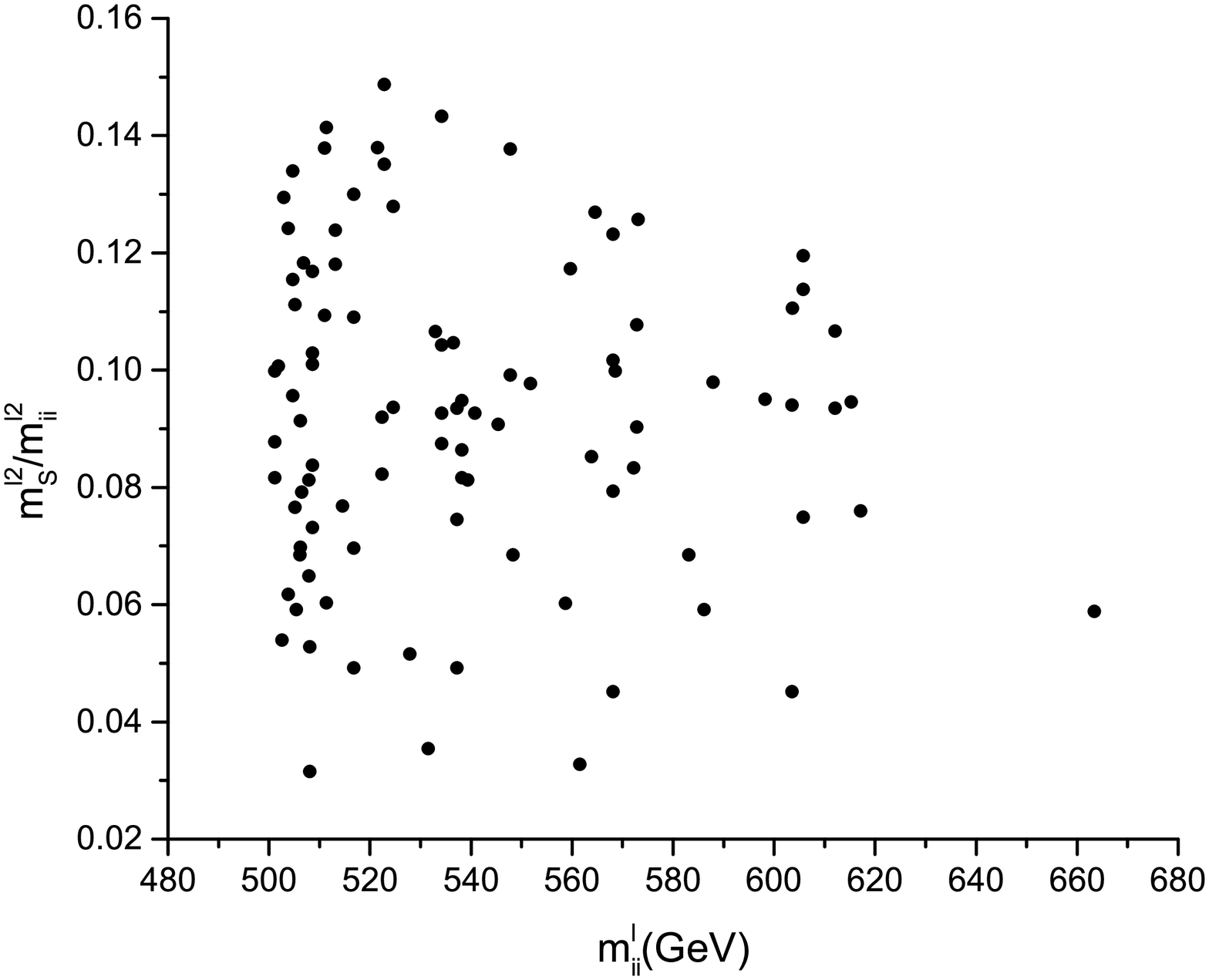}
\caption{Neutrino mass with correct hierarchy in the $m^l_{ii}$ - $m^l_s$ plane}
\label{mliimls}
\end{figure}

The scanning processes are divided into three steps. First of all, scan from parameter space
\begin{eqnarray}
&&0GeV \le M_1<600GeV,~~~320GeV \le M_2<600GeV,~~~800GeV \le M_3 \le 2000GeV \nonumber \\
&&1 \le \tan{\beta} \le 10,~~~0.1 \le \lambda  \le 0.7,~~~0.1 \le \kappa \le 0.7,~~~100GeV \le \mu_{eff} \le 1000GeV  \nonumber \\
&&-5000GeV \le A_{\lambda} \le 5000GeV,~~~-5000GeV \le A_{\kappa} \le 5000GeV . \nonumber \\
&& (500GeV)^2 \le m^{l2}_{ii}=m^{E2}_{ii} \le (1500GeV)^2
\end{eqnarray}
During this step, NMSSMTools is hardly modified except  the lower-bound of the lightest Higgs mass. The Higgs mass bound is temporarily set as $112.7GeV < M_{Higgs} < 128.7GeV$. Just because we opened the anomalous $g_{\mu}-2$ constraints, we can see from Figure \ref{mliimls} that $m^l_{ii}$ concentrate below 600GeV. 

Then the second step to calculate the modification of the Higgs mass by the right-handed neutrino sectors is applied. We modified NMSSMTools-4.2.1 by changing the part of the Higgs sectors considering the effects from the right-handed neutrino sectors. Here, the Higgs mass bound is set back to $122.7GeV<M_{Higgs}<128.7GeV$ in order to filter the consequences output from the previous step.

The final and the most important step is to decide the remaining parameters inside the range
\begin{eqnarray}
&& -1 \times 10^{-6} \le y_{N1} \le 1 \times 10^{-6}, ~~~ -1 \times 10^{-6} \le y_{N2} \le 1 \times 10^{-6}, ~~~ -1 \times 10^{-6} \le y_{N3} \le 1 \times 10^{-6}, \nonumber \\
&& -300TeV \le A_{yN1} \le 300TeV,~~~-300TeV \le A_{yN2} \le 300TeV,~~~-300TeV \le A_{yN3} \le 300TeV,~~~ \nonumber \\
&& |m^{l2}_{ij}| \le \frac{1}{10}m^{l2}_{ii}(\text{for each $i \neq j$})
\end{eqnarray}
for each of the parameter point obtained from the last step. We scanned randomly in this area at first, and tested each point to see whether it can lead to the correct neutrino mass-squared difference, then start from the nearest point to ``jog'' near the correct position
\begin{eqnarray}
7.12 \times 10^{-5}<\Delta m_{21}^2 < 8.20 \times 10^{-5} \nonumber \\
2.31 \times 10^{-3}<|\Delta m_{31}^2| < 2.74 \times 10^{-3}.
\end{eqnarray}
This process takes most of the time.

If we rotate the basis of $L_i$ by a unitary matrix $V$, the parameters $y_{Ni}$, $y_{Ni} A_{yNi}$(i is not summed up) correspondingly behave like a ``vector-like representation'' of $V$, so define the scalar-like norm of these two parameters
\begin{eqnarray}
&& y_{NS} = \sqrt{\sum_{i=1}^3 {y_{Ni}^2}}, \nonumber \\
&& A_{yNS} y_{NS} = \sqrt{\sum_{i=1}^3 {(A_{yNi} y_{Ni})^2)}}.
\end{eqnarray}
Though $m^{l2}_{ij}(i \ne j)$ do not transform like a vector, we still define a ``scalar-like''
\begin{eqnarray}
m^{l2}_s = \sqrt{(m^{l2}_{12})^2+(m^{l2}_{13})^2+(m^{l 2}_{23})^2}.
\end{eqnarray}

\begin{figure}[center]
\includegraphics[width=4in]{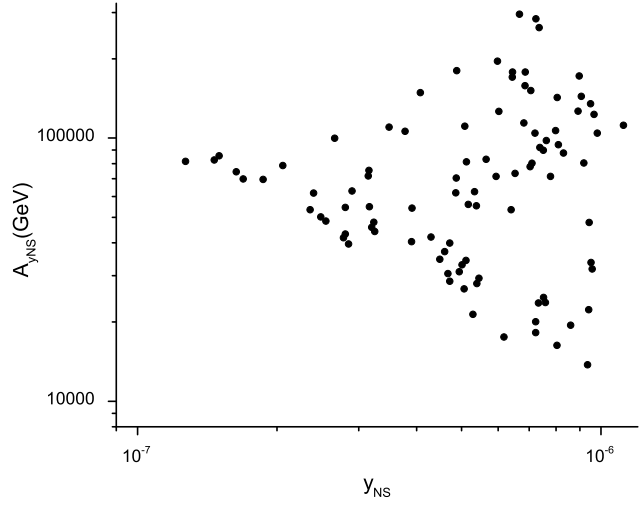}
\caption{Neutrino mass with correct hierarchy in the $A_{yNS}$ - $y_{NS}$ plane} \label{yNAyN}
\end{figure}
\begin{figure}[center]
\includegraphics[width=4in]{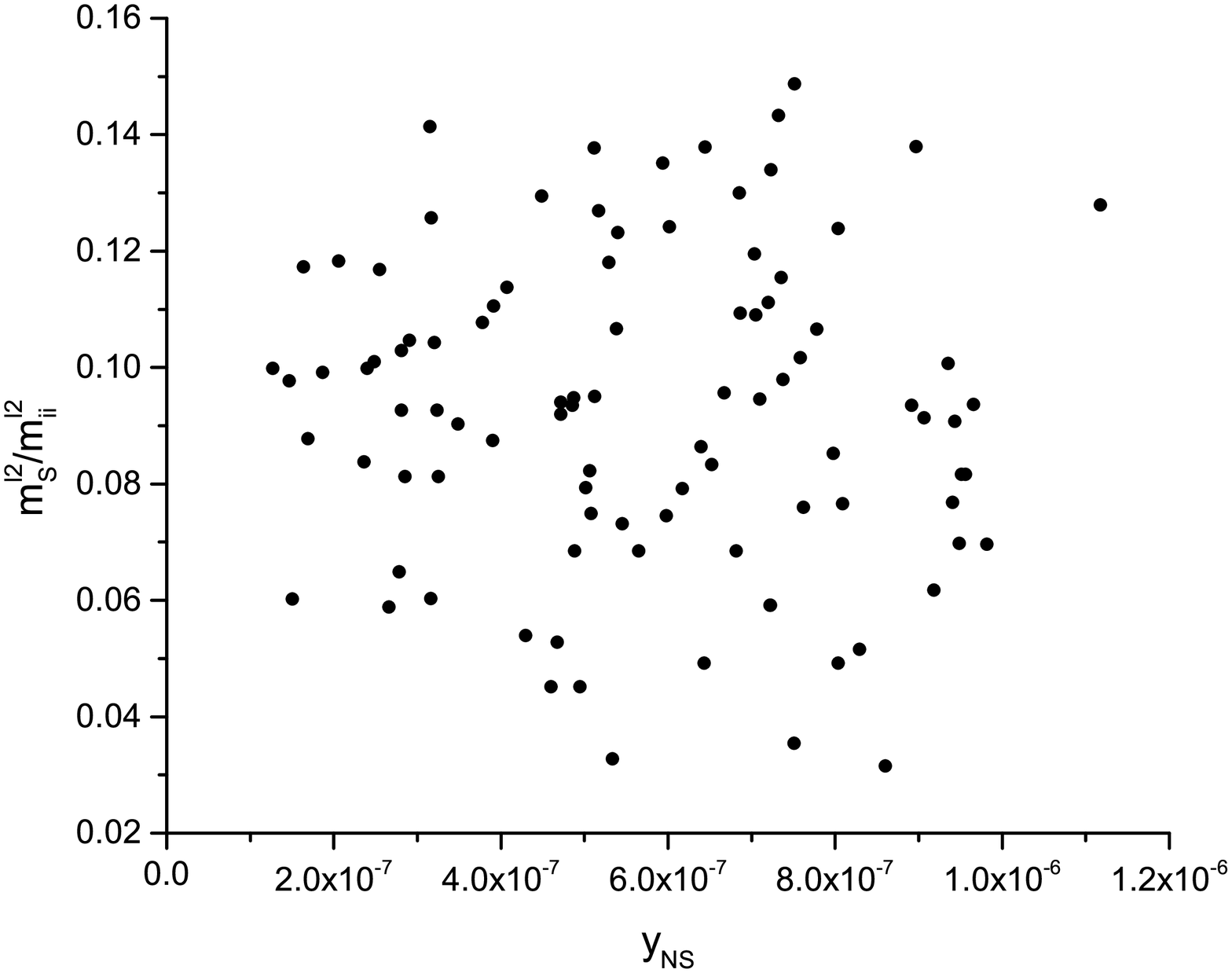}
\caption{Neutrino mass with correct hierarchy in the $y_{NS}$ - $m^l_s$ plane}
\label{yNmls}
\end{figure}
\begin{figure}[center]
\includegraphics[width=4in]{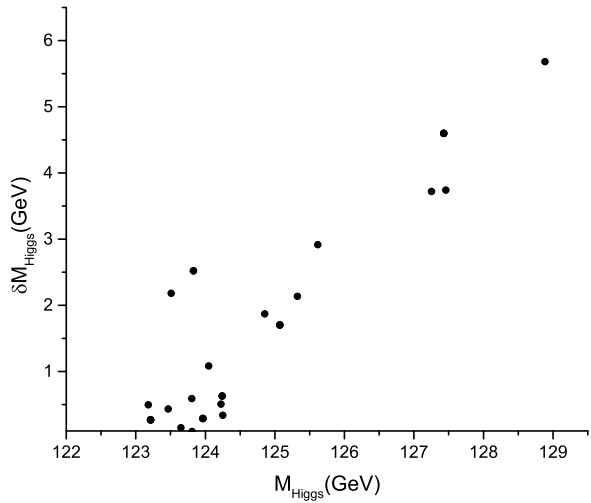}
\caption{The extra higgs mass $\delta M_{Higgs}$ contributed by right-handed neutrino}
\label{mhiggs}
\end{figure}

Figure \ref{yNAyN}, \ref{yNmls} and \ref{mliimls} show the parameter points from the aspects of $y_{NS}$ - $A_{yNS}$, $y_{NS}$ - $\frac{m^{l2}_s}{m^{l2}_{ii}}$, $m^l_{ii}$ - $\frac{m^{l2}_s}{m^{l2}_{ii}}$ planes separately, where $m^{l}_{ii}=\sqrt{m^{l2}_{ii}}$ denotes the soft mass of the sneutrino. Especially from the Figure \ref{yNAyN}, we can confirm our previous discussion that relatively large $A_{yN} \sim 10^2 \text{TeV}$ are needed in order to generate a relatively large loop contribution to the originally massless neutrinos in tree-level. However, $A_{yN} y_N \sim 1\text{GeV}$, which is so small that they can be ignored in most of the phenomenology analysis. One might believe that a large A-term could break the stability of the correct vacuum, resulting in some other deeper vacua, as mentioned in the Ref. \cite{MultiRightHandedNMSSM}. However, in this situation, we should also consider the gauge quartic terms such as $\frac{1}{8}(g^2 + g^{\prime2})(|H_u^0|^2-|H_d^0|^2)^2$, and the $\lambda_N^2 N^2 N^{*2}$ from other yukawa terms. After a some calculation, we can see these quartic terms actually stabilize the correct vacuum. In fact, when we calculate the contribution to higgs mass from the right-handed neutrino sectors, we omitted these A-terms. By the way, Figure \ref{mhiggs} shows that Higgs mass does have the possibility to recieve a relatively large correction, which is compatible with that in \cite{LotsofMistakes}.

\section{Phenomenology}

Although we would like the $A_{yN}$ terms to be large enough in order for the loop corrections to be comparable with the tree-level terms, loop-corrections still tend to be smaller. They only give smaller masses to the two neutrinos which are originally massless up to tree-level see-saw mechanism. Thus, we predict a normal-hierarchy mass spectrum of neutrinos without any degeneracy.

We have mentioned that we need off-diagonal terms in the soft mass-squared terms of sleptons, which may lead to visible $\mu \rightarrow e + \gamma$ decay. The branching ratio of this chain is estimated in \cite{primer}
\begin{eqnarray}
Br(\mu \rightarrow e \gamma) \sim \left( \frac{m^{l2}_{e\mu}}{m^{l2}_{ii}}\right)^2 \left( \frac{100GeV}{m^l_{ii}} \right)^4 10^{-6}.
\end{eqnarray}
From the data showed in Figure \ref{mliimls}, we can estimate that $Br(\mu \rightarrow e + \gamma) $ varies from $10^{-12}~~-~~10^{-10}$. Compared with the PDG data $Br(\mu \rightarrow e + \gamma) < 2.4 \times 10^{-12}$\cite{PDG}, it means at least some of the points have passed the constraint and some are on the edge of the experimental bound. As mentioned before, if we assume that the source of muon anomalous $g_\mu-2$ comes from the NMSSM sectors, the soft masses of sleptons are strictly constrained to be much smaller than $1 \text{TeV}$, and we imposed this constraint during our scanning procedure. If we release such a constraint, the slepton mass can reach above $1\text{TeV}$ so that $\mu \rightarrow e + \gamma$ decay is totally invisible.

In Ref. \cite{PDG}, bounds on other muon FCNC decay channels are listed, such as $\mu \rightarrow 3e$, $\mu \rightarrow e+2\gamma$. These bounds on branching ratios are roughly of the similar order of magnitude of $\mu \rightarrow e + \gamma$, but their Feynmann-diagrams usually involve a higher-order perturbative expansion, thus the effects are depressed, so we did not discuss them. On the other hand, the $\tau$-FCNC bounds are much looser, so we did not talk about them either.

As mentioned in Appendix I, TeV scale right-handed neutrinos have no hope to become dark matter, as they usually decay quickly. If R-parity is conserved, the scalar partner of $N$ might become the LSP, thus the candidate of the dark matter, and the corresponding phenomenology is discussed in \cite{SNeutrino_Dark_Matter}. As we have noted, we added the effects of the right-handed scalar neutrinos in our theory when calculating the dark matter relic density.

Generally, the possibility to discover a right-handed neutrino directly on a collider is significantly suppressed due to the rather small Yukawa coupling $y_N$ in the case of TeV see-saw mechanism, unless other physical sectors beyond the Standard Model which appear to interact with right-handed neutrinos exist\cite{LHC_Phenomenology}. In the circumstances of NMSSM, the interaction between the right-handed neutrino and the singlet Higgs (characterized by the magnitude of $y_N$ and $A_{yN}$) can be relatively large. If we are able to observe the singlet Higgs directly in the future, we can take a glimpse of right-handed neutrinos by watching the properties of the singlet, e.g. an invisible decay chain in the case that the right-handed (s)neutrino is lighter than the singlet higgs, or its correction to the propagator when it is heavier.

\section{Conclusion}

In this article, we have shown that NMSSM with a $Z_3$ symmetry extended with only one right-handed neutrino superfield can generate a complete spectrum of three massive light Majorana left-handed neutrinos. The tree-level see-saw mechanism can only generate one massive neutrino, with the remaining two acquiring masses from radiative one-loop corrections. To accumulate the loop effects in order it can be comparable with the tree-level in quantity, we need relatively large $A_{yN} \sim 10^2 TeV$, however other phenomenological effects from them are suppressed by the Yukawa $y_N \sim 10^{-7}$. Though off-diagonal terms are needed in the soft mass-square terms of the sleptons, we are still able to acquire the correct neutrino mass differences without conflicting with the phenomenology constraints. We also showed that once the correct mass-difference is acquired, any figure of mixing angles is allowed, and of course so is the one measured by experiments. We also confirmed that the right-handed neutrino can contribute to Higgs mass by its coupling with the Higgs sectors.

\begin{acknowledgments}

We would like to thank Professor Chun Liu, Dr.~Zhen-hua Zhao, Mr.~Ye-ling Zhou for helpful discussions.
This work was supported in part by the National Natural Science Foundation of China under Nos. 11375248, and by the
National Basic Research Program of China under Grant No. 2010CB833000.

\end{acknowledgments}

\newpage

\appendix*

\section{I}

In order to consider the effects of $N$, $\tilde{N}$ during the calculation of dark matter decay modes, we should give all of the corresponding vertices.
\begin{eqnarray}
&&V_{h_i \tilde{N}_R \tilde{N}_R }=\sqrt{2}\lambda_{N}\lambda\left(v_{u}S_{i2}+v_{d}S_{i1}\right)
- \sqrt{2}\left(2\lambda_{N}\kappa v_{s} +4\lambda_{N}^{2}v_{s}+\lambda_{N}A_{N}\right)S_{i3}, \label{vertices_start}\\
&& V_{h_i \tilde{N}_I \tilde{N}_I }=-\sqrt{2}\lambda_{N}\lambda\left(v_{u}S_{i2}+v_{d}S_{i1}\right)
+ \sqrt{2} \left(2\lambda_{N}\kappa v_{s} -4\lambda_{N}^{2}v_{s}+\lambda_{N}A_{N}\right)S_{j3}\\
&&V_{h_i h_j \tilde{N}_R \tilde{N}_R }=-\lambda_{N}\left[2\kappa S_{j3}S_{i3}-\lambda(S_{j1}S_{i2}+S_{i1}S_{j2})\right]-4\lambda_{N}^{2}S_{j3}S_{i3}\\
&&V_{h_ih_j \tilde{N}_I \tilde{N}_I }=\lambda_{N}\left[2\kappa S_{j3}S_{i3}-\lambda(S_{j1}S_{i2}+S_{i1}S_{j2})\right]-4\lambda_{N}^{2}S_{j3}S_{i3}\\
&&V_{a_i \tilde{N}_R \tilde{N}_I } = -2\lambda_N(-\lambda v P_{i1}/\sqrt2 +\sqrt2 \kappa v_s P_{i1}) + \sqrt2\lambda_N A_N P_{i2},\\
&&V_{a_i a_j \tilde{N}_R \tilde{N}_R} = 2\lambda_N(-\lambda \sin\beta\cos\beta
P_{i1}P_{j1}+\kappa P_{i2}P_{j2})-4\lambda_N^2P_{i2}P_{j2},\\
&&V_{a_i a_j \tilde{N}_I \tilde{N}_I} = -2\lambda_N(-\lambda \sin\beta\cos\beta P_{i1}P_{j1}+\kappa P_{i2}P_{j2})-4\lambda_N^2P_{i2}P_{j2},\\
&& V_{h_i NN}=-\sqrt{2}\lambda_NS_{i3}~~~~~~~~
V_{a_i NN}=\sqrt{2}i\lambda_N P_{i2}\gamma^{5},\\
&&V_{\chi_i \tilde{N}_R N}=-\lambda_N\frac{N_{i5}}{2\sqrt{2}}~~~~~~~~~~ V_{\chi_i \tilde{N}_I N}=\lambda_N\frac{iN_{i5}\gamma^{5}}{2\sqrt{2}}, \\
&& V_{\tilde{N}_R \tilde{N}_R H^+ H^-} = -\lambda_N \lambda \cos{\beta} \sin{\beta}, \\
&& V_{\tilde{N}_I \tilde{N}_I H^+ H^-} = \lambda_N \lambda \cos{\beta} \sin{\beta}, \\
&& V_{\tilde{N}_R \tilde{N}_R \tilde{N}_R \tilde{N}_R} = V_{\tilde{N}_I \tilde{N}_I \tilde{N}_I \tilde{N}_I}=6 \lambda_N^2 , \\
&& V_{\tilde{N}_R \tilde{N}_R \tilde{N}_I \tilde{N}_I}=2 \lambda_N^2 \label{vertices_end},
\end{eqnarray}
where the definition of the diagonalized field $h_i$, $a_i$, together with their diagonalizing matrix $S_{ij}$, $P_{ij}$ is similar to the tree-level ones in Appendix II. However, unlike appendix II, we should note that when applying these vertices to calculate the dark matter relic density, we should use the renormalized version of $h_i$, $a_i$, $S_{ij}$, $P_{ij}$. Part of the vertices listed here is copied and modified from \cite{LotsofMistakes}.

All of them are input into the MicroOMEGAS\cite{MicroOMEGAS} model files inside the NMSSMTools, and most of the vertices will also be used when calculating the Higgs mass. Because $\tilde{N}$ is assigned with the odd R-parity, $\tilde{N}$ rather than fermionic $N$ is set as the candidate of the dark matter. One may ask the question that whether right-handed neutrinos can play the role of dark matter if they decay slowly enough. According to \cite{RHN_Dark1}\cite{RHN_Dark2}, right-handed neutrinos heavier than 1 GeV usually decay less than one second, so it is impossible for them to become the dark matter.

\section{II}

The tree-level Higgs mixing matrix should be calculated before proceeding the renormalization scheme. Define $h^{bare} = [Re(H_u^0), Re(H_d^0), S_R]$, the CP-even mass-eigenstate higgs in tree level  are
\begin{eqnarray}
h_i = S_{ij} h_j^{bare}
\end{eqnarray}
where $S_{ij}$ is an orthogonal rotation matrix. For $[Im(H_u^0), Im(H_d^0), S_I]$, define
\begin{eqnarray}
A=\cos{\beta} Im(H_u^0) + \sin{\beta} Im(H_d^0) \nonumber \\
G=-\sin{\beta} Im(H_u^0) + \cos{\beta} Im(H_d^0),
\end{eqnarray}
then drop the Goldstone state $G$, and diagonalize $(A,S_I)$ into
\begin{eqnarray}
a_1 = P_{11} A + P_{12} S_I \nonumber \\
a_2 = P_{21} A + P_{22} S_I,
\end{eqnarray}
we acquire two CP-odd mass-eigenstates. 
To diagonalize the neutralino mass matrix $\mathcal{M}_N$ in the basis $\psi^0 = (-i \lambda_1, -i \lambda_2, \psi^0_u, \psi^0_d, \psi_s)$, define $\chi^0_i = N_{ij} \psi_j^0$.

To calculate the contribution to the higgs mass from the right-handed neutrino, we need to choose a renormalization scheme. We choose the parameter set
\begin{eqnarray}
M_Z, M_W, e, \underbrace{ t_{H_u}, t_{H_d}, t_{H_s}}_{\text{on-shell sheme}}, \underbrace{M_{H^\pm} \tan{\beta}, \lambda, v_s, \kappa, A_\kappa}_{\text{$\overline{DR}$ scheme}},
\end{eqnarray}
where $t_{H_u}$, $t_{H_d}$, $t_{H_s}$ are the tadpoles of the CP-even Higgs fields. $M_Z$, $M_W$, $e$ need not be renormalized and are just regarded as SM input parameters. Replace the parameters by the renormalized ones plus the counterterms:
\begin{eqnarray}
\begin{array}{ccc}
t_{H_u} \rightarrow t_{H_u} + \delta t_{H_u}, & t_{H_d} \rightarrow t_{H_d} + \delta t_{H_d}, & t_{H_s} \rightarrow t_{H_s} + \delta t_{H_s},  \\
\tan{\beta} \rightarrow \tan{\beta} + \delta \tan{\beta}, & \lambda \rightarrow \lambda + \delta \lambda, & \kappa \rightarrow \kappa + \delta \kappa \\
v_s \rightarrow v_s + \delta v_s, & A_\kappa \rightarrow A_\kappa + \delta A_\kappa, & M_{H^\pm}^2 \rightarrow M_{H^\pm}^2 + \delta M_{H^\pm}^2
\end{array} \label{Counter_Terms}
\end{eqnarray}

renormalized two-point functions need to be calculated in mass-eigenstate basis $(H_i),(i=1-3)$, by using the vertices listed in (\ref{vertices_start}-\ref{vertices_end}), and then to be rotated into the original basis $(H_u, H_d, S)$. The field-renormalization constant $\delta Z_{H_i H_i}$ is calculated through
\begin{eqnarray}
\delta Z_{H_i H_i} = - \left.\frac{\partial \Sigma_{H_i H_i}}{\partial k^2}\right|^{\text{div}}_{k^2 = (M_{H_i}^{(0)})^2}.
\end{eqnarray}
To get $\delta Z_{H_u}$, $\delta Z_{H_d}$, $\delta Z_S$, equations
\begin{eqnarray}
\delta Z_{H_i H_i} = |S_{i1}|^2 \delta Z_{H_d} + |S_{i2}|^2 \delta Z_{Hu} + |S_{i3}|^2 \delta Z_S \text{\qquad} (i=1,2,3)
\end{eqnarray}
should be solved. We also need to calculate $\Sigma_{A_i A_j} (k^2)$, in order to extract some divergent terms. These constants determine the counterterms listed in (\ref{Counter_Terms}), and we list them in the following text.
\begin{eqnarray}
\delta t_{H_i} = S_{ji} t_{h_j}^{(1)}\text{\qquad}(i=u,d,s,\text{\quad}j=1,2,3),
\end{eqnarray}
where $t_{h_j}^{(1)}$ denote the one-loop Higgs tadpoles.
\begin{eqnarray}
\delta \tan{\beta} &=& \left[ \frac{\tan{\beta}}{2} ( \delta Z_{H_u} - \delta Z_{H_d} ) \right]_{div} , \\
\delta \lambda &=& \frac{e^2}{4 \lambda M_W^2 s_W^2} \left[ \Sigma_{P,11} (M_{P,11}^2) \right]_{div},
\end{eqnarray}
where $\Sigma_{P,11} = P_{i1} \Sigma_{A_i A_j} P_{j1}$.
\begin{eqnarray}
\delta M_{H^\pm}^2 &=& \left. Re(\Sigma_{H^{\pm} H^{\pm}} (M_{H^\pm}^2)) \right|_{div}\\
\delta v_s &=& \left[ -v_s \frac{\delta \lambda}{\lambda} - \delta M_{H^\pm}^2 \right]_{div} \\
\delta \kappa &=& \frac{1}{2 v_s} \delta (\mathcal{M})_{SS} - \kappa \frac{\delta v_s}{v_s} \\
\delta A_\kappa &=&  \left[ -\frac{1}{3 \kappa v_s} [\Sigma_{P,22} (M_{P,22}^2) - \delta f ] - A_\kappa [ \frac{\delta \kappa}{\kappa} + \frac{\delta v_s}{v_s}] \right]_{div},
\end{eqnarray}
where
\begin{eqnarray}
\delta f &=& \frac{t_{H_S}}{\sqrt{2} v_s} - \frac{ M_W \sin{\theta_W} s_{\beta} c_{\beta}^2 c_{\beta_B}^2}{e v_s^2 c_{\Delta \beta}^2} (t_{H_u} + t_{H_d} t_{\beta} t_{\beta_B}^2 ) \\
&+& \frac{M_W^2 s_W^2 s_{2 \beta}^2}{2 e^2 v_s^2 c_{\Delta \beta}^2} ( M_{H^\pm}^2 - M_W^2 c_{\Delta \beta}^2 )
+ \frac{\lambda M_W^2 \sin{\theta_W}^2 s_{2 \beta} }{2 e^4 v_s^2} ( 2 \lambda M_W^2 \sin{\theta_W^2} s_{2 \beta} + 6 \kappa e^2 v_s^2) \nonumber,
\end{eqnarray}
and $\beta_B$ denotes the tree-level $\beta$.

After the determination of the counter-terms, the Higgs mass sectors are differentiated and the related terms are replaced with the counter terms acquired through the previous steps. The elements of the mass matrix of the Higgs sectors are listed below:
\begin{eqnarray} 
\nonumber M^2_{S_{11}} &=& \frac{e c_{\beta} c_{\beta_B}}{2 M_W s_W c^2_{\Delta\beta}}  [-t_{H_d} s_{\beta_B} t_{\beta_B} + t_{H_u} s_{\beta_B} (t_{\beta} t_{\beta_B} + 2)] \\ 
&+& \frac{c_\beta^2 }{c^2_{\Delta\beta}}[M_{H^\pm}^2+(M_Z^2 t_\beta^2 -M_W^2) c^2_{\Delta\beta}] + \frac{2 \lambda^2 M_W^2 s_W^2 c^2_{\beta}}{e^2}, \\
M^2_{S_{12}} &=& \frac{e c_\beta c^2_{\beta_B} }{2 M_W  s_W c^2_{\Delta\beta}} [t_{H_d} t_{\beta} t^2_{\beta_B} + t_{H_u}] - \frac{s_{\beta} c_{\beta} }{c^2_{\Delta\beta}} [ M_{H^\pm}^2 +(M_Z^2-M_W^2) c^2_{\Delta\beta} ] + \frac{\lambda^2 M_W^2 s_W^2 s_{2 \beta}}{e^2}, \\ \nonumber
M^2_{S_{13}} &=& \frac{c^2_{\beta} c^2_{\beta_B} }{\sqrt{2}v_s c^2_{\Delta\beta}} [ t_{H_d} t_{\beta} t^2_{\beta_B} +  t_{H_u} ] + \frac{\sqrt{2} M_W s_W s_\beta c^2_\beta }{e v_s c^2_{\Delta\beta}} [M_W^2
c^2_{\Delta\beta} -M_{H^\pm}^2] \\
&+& \frac{\sqrt{2}\lambda M_W s_W c_\beta v_s }{e} [2 \lambda t_\beta - \kappa ] + \frac{-2\sqrt{2} \lambda^2 M_W^3 s_W^3 s_\beta c^2_{\beta}}{e^3 v_s}, \\
M^2_{S_{22}} &=& \frac{e c_\beta c^2_{\beta_B} }{2 M_W s_W c^2_{\Delta\beta}} [t_{H_d} (2 t_\beta t_{\beta_B} +1) -t_{H_u} t_\beta] \nonumber \\
&+& \frac{s^2_{\beta} }{c^2_{\Delta\beta}}[M_{H^\pm}^2+(M_Z^2 t^{-2}_\beta-M_W^2) c^2_{\Delta\beta}] + \frac{2 \lambda^2 M_W^2 s_W^2s^2_\beta}{e^2}, \\
M^2_{S_{23}} &=& \frac{s_\beta c_\beta c^2_{\beta_B} }{\sqrt{2} v_s
  c^2_{\Delta\beta}} [t_{H_d} t_{\beta} t^2_{\beta_B} + t_{H_u}] + \frac{\sqrt{2} M_W s_W s_\beta^2 c_\beta }{e v_s c^2_{\Delta\beta}} [M_W^2 c^2_{\Delta\beta} -M_{H^\pm}^2] \\
&+& \frac{\sqrt{2}\lambda M_W s_W c_\beta v_s }{e} [2 \lambda - \kappa t_\beta] + \frac{-2\sqrt{2} \lambda^2 M_W^3 s_W^3 s^2_{\beta} c_{\beta}}{e^3 v_s} \\
\nonumber M^2_{S_{33}} &=&  \nonumber \kappa A_\kappa v_s +4 \kappa^2 v_s^2+\frac{t_{H_s}}{\sqrt{2}v_s} + \frac{M_W s_W s_\beta c_\beta^2 }{e^2 v_s^2 c^2_{\Delta\beta}} [2 M_{H^\pm}^2 M_W s_W s_\beta - e (t_{H_d}  t_\beta s^2_{\beta_B} + t_{H_u} c^2_{\beta_B})] \\
&+& \frac{M_W^2 s_W^2 s_{2\beta}}{2 e^4 v_s^2}[2\lambda^2 M_W^2 s_W^2 s_{2\beta} - 2\kappa \lambda e^2 v_s^2 - M_W^2 e^2 s_{2\beta}]. \\
M^2_{P_{11}} &=& \frac{2 \lambda^2 M_W^2 s_W^2 c^2_{\Delta \beta}}{e^2} +M_{H^\pm}^2 -M_W^2 c^2_{\Delta \beta},\\
\nonumber M^2_{P_{12}} &=& \frac{M_W  s_W s_{2 \beta}} {\sqrt{2} e v_s c_{\Delta \beta}} [M_{H^\pm}^2 - M_W^2 c_{\Delta\beta}^2] -\frac{c_{\beta} c^2_{\beta_B}}{\sqrt{2} v_s c_{\Delta \beta}} [t_{H_u} + t_{H_d} t_{\beta} t^2_{\beta_B}] \\
&+& \frac{\lambda M_W s_W c_{\Delta \beta}}{\sqrt{2} e^3 v_s} [2\lambda M_W^2 s_W^2 s_{2\beta} - 6\kappa e^2 v_s^2], \\
M^2_{P_{13}} &=& M_{H^\pm}^2 t_{\Delta \beta} +\frac{M_W^2 s_{2 \Delta \beta}}{2e^2} [2 \lambda^2 s_W^2 - e^2] +\frac{e c_{\beta_B}}{2 M_W s_W c_{\Delta \beta}} [t_{H_d} t_{\beta_B} -t_{H_u}], \\
\nonumber M^2_{P_{22}} &=&-3 A_\kappa \kappa v_s + \frac{t_{H_s}}{\sqrt{2} v_s} - \frac{M_W s_W s_\beta c_\beta^2 c^2_{\beta_B}}{e v_s^2 c_{\Delta \beta}^2}[t_{H_u}+ t_{H_d} t_\beta t^2_{\beta_B}] \\
&+& \frac{M_W^2 s_W^2 s_{2\beta}^2}{2 e^2 v_s^2 c_{\Delta\beta}^2} [M_{H^\pm}^2-M_W^2 c_{\Delta \beta}^2] + \frac{\lambda M^2_W s^2_W s_{2\beta}}{e^4 v_s^2} [\lambda M_W^2 s_W^2 s_{2\beta} + 3\kappa e^2 v_s^2], \\ \nonumber
M^2_{P_{23}} &=& \frac{M_W s_W s_{2\beta}} {2\sqrt{2}ev_s c_{\Delta \beta}} [2 M_{H^\pm}^2 t_{\Delta \beta} - M_W^2 s_{2\Delta \beta}] -\frac{c_{\beta} c^2_{\beta_B} t_{\Delta \beta} }{\sqrt{2}v_s c_{\Delta \beta}} [t_{H_u} + t_{H_d} t_{\beta} t^2_{\beta_B}] \\
&+& \frac{\lambda M_W s_W s_{\Delta \beta}}{\sqrt{2} e^3 v_s} [2\lambda M_W^2 s_W^2 s_{2\beta} - 6\kappa e^2 v_s^2],\\ 
\nonumber M^2_{P_{33}} &=& M_{H^\pm}^2 \tan ^2{\Delta \beta}  + \frac{M_W^2 \sin^2{\Delta \beta}}{e^2} [2 \lambda^2 s_W^2 - e^2] \\
&+& \frac{e }{2 M_W s_W c^2_{\Delta \beta}} [t_{H_d} c_{\beta - 2 \beta_B} -t_{H_u} s_{\beta - 2 \beta_B}],
\end{eqnarray}
where $c_X$, $s_X$, $t_X$ denote respectively $\cos{X} $, $\sin{X}$, $\tan{X}$. The $M_{S_ij}^2$ are mass-square matrix elements in the basis $(Re(H_u^0), Re(H_d^0), S_R)$, and $M_{P_ij}^2$ are the elements in the basis $(A,S_I,G)$.

Theoretically, all divergent $\frac{1}{\epsilon}$ should be precisely cancelled after the renormalization scheme in the final results. We checked this carefully. Though the existence of matrices $P_{ij}$, $S_{ij}$ blurred the final expressions, divergent terms proportional to $\frac{1}{\epsilon}$ must be independent of field basis, so we can directly calculate the divergent part by setting $P_{ij}=S_{ij}=\delta_{ij}$, which is much easier to operate. We checked and modified the formulae listed in \cite{LotsofMistakes} by verifying whether the infinite parts of the diagrams can be automatically cancelled. Only when confirming this, can we calculate on.

\end{document}